\documentclass[12pt,a4paper]{article}

\ifx\pdfoutput\undefined
\usepackage[usenames]{color} 
\usepackage[dvips]{graphicx}
\else
\usepackage{color} 
\definecolor{BrickRed}{rgb}{0.85,0.15,0.25}
\definecolor{MidnightBlue}{rgb}{0,0.45,0.85}
\definecolor{ForestGreen}{rgb}{0,0.85,0.45}
\usepackage{graphicx}
\usepackage{epstopdf}
\fi
\usepackage{subfigure}
\usepackage{umoline}
\usepackage{latexsym,amsmath,amssymb}
\allowdisplaybreaks
\renewcommand{\thefootnote}{\fnsymbol{footnote}}
\newcommand{\acknowledgments}{
$\\${\bf Acknowledgments}\newline}

\begin{document}
\begin{center}
{{{\Large {\bf Note on uncertainty relations in doubly special relativity and rainbow gravity}}}}\\[10mm]
{Edwin J. Son$^{a}$\footnote{email:eddy@nims.re.kr} and
Wontae Kim$^{b}$\footnote{email:wtkim@sogang.ac.kr}}\\[8mm]

{{${}^{a}$ Division of Computational Sciences in Mathematics,\\
National Institute for Mathematical Sciences, Daejeon 305-390, Republic of Korea}\\[0pt]
${}^{b}$ Department of Physics, Sogang University, Seoul, 121-742, Republic of Korea\\[0pt]
}
\end{center}
\vspace{2mm}
\begin{abstract}
We present commutation relations depending on the rainbow functions
which are slightly different from the well-known results.
However, the advantage of these new commutation relations are
compatible with the calculation of the Hawking temperature
in the rainbow Schwarzschild black hole.

\end{abstract}
\vspace{5mm}



\vspace{1.5cm}

\hspace{11.5cm}{Typeset Using \LaTeX}
\newpage
\renewcommand{\thefootnote}{\arabic{footnote}}
\setcounter{footnote}{0}

\newcommand{\lp}{\ell_{P}}


One of the most interesting issues in quantum theory of gravity
is that there appears a minimal length
from various points of view such as
string theory~\cite{Amati:1987wq,Kostelecky:1988zi}, loop quantum
gravity~\cite{Gambini:1998it}, and Gedanken
experiments~\cite{Garay:1994en,Mead:1964zz,Calmet:2004mp}.
The minimal length is related to the modification of quantum-commutation
relations~\cite{Maggiore:1993rv,Maggiore:1993kv,Maggiore:1993zu,Das:2008kaa},
so that the Heisenberg uncertainty principle is promoted to
the generalized uncertainty principle (for a recent extensive review, see
Ref.~\cite{Tawfik:2014zca}).
Meanwhile,
it has also been claimed that a modified dispersion relation could come
from the deformation of
classical relativity so called the doubly special
relativity~\cite{AmelinoCamelia:2008qg,
AmelinoCamelia:2000ge,AmelinoCamelia:2000mn,AmelinoCamelia:2002vy}
which is the extended version for the special theory of relativity
in the sense that both the Plank-scale energy and the speed of light are
required to be invariant under any inertial frames throughout the nonlinear realization
of the Lorentz transformations on the momentum space.
A simple realization of the modified dispersion relation for the doubly special
relativity was exhibited in the flat spacetime~\cite{Magueijo:2001cr, Magueijo:2002am}
and the curved spacetime~\cite{Magueijo:2002xx}.

Now, it is worth noting that the uncertainty principle from quantum mechanics is
indeed independent of the dispersion relation from the classical
relativity. So,
it is not until a commutation relation or uncertainty relation
are imposed that
one can relate the energy scale defined in the momentum
space with a length scale defined in the spacetime coordinates.
In this respect, quantum commutation relations
for a specific nonlinear realization
could be obtained by using the modified generators satisfying
the conventional Lorentz algebra in Ref.~\cite{Magueijo:2002am}.
Then, for more general nonlinear realizations,
the commutation relations to implement the doubly special relativity at the level
of quantum mechanics were obtained
by introducing space time coordinates as the generators of translations
in the linearly transforming energy-momentum variables~\cite{Cortes:2004qn}.

On the other hand, one can obtain the black hole temperature by using
the uncertainty principle since the emitted particle energy from the black hole
is related to the uncertainty relation.
For the usual Schwarzschild black hole,
the position uncertainty of the emitted particles
is assumed to be the radius of the event horizon of $\Delta x \sim r_H$, so that
the momentum uncertainty in the spherically symmetric geometry
can be obtained as $p \sim \Delta p \sim 1/r_H$
in virtue of the usual
Heisenberg uncertainty relation.
Next, using the standard dispersion relation as $E=p$ for the massless case,
the Hawking temperature which is proportional to the energy of the
emitted particles can be estimated with a calibration factor as
$T_H =1/(4\pi r_H)$.
This heuristic derivation was illuminated in the calculation of the
temperature of the black hole subject to the generalized uncertainty principle~\cite{Adler:2001vs}.
The key ingredient is that the Hawking temperature
is connected with the uncertainty relation for a given dispersion relation, so that
one can get various expressions for the temperature depending on commutator relations,
which have been used in the thermodynamic analysis of rainbow black holes~\cite{Ali:2014xqa,Gim:2014ira}.

In this work, we would like to find quantum-mechanical commutation relations
depending on the rainbow functions
and identify the relevant uncertainty relation
to the calculation of the Hawking temperature.
It will be shown that the uncertainty relation between the position and
the momentum 
can be independent of the rainbow functions, so that
it gives the same Hawking temperature as that
of calculation of
the surface gravity in the rainbow Schwarzschild black hole.

Considering a transformation implemented by $U(p_0)$ acting on $p_\mu$ as $U(p_0) \cdot p_\mu = (p_0 f(p_0), p_i g(p_0))$,
the modified dispersion relation can be written as~\cite{Magueijo:2001cr,Magueijo:2002am}
\begin{equation}
  \label{mdr}
  \eta^{\mu\nu} (U(p_0) \cdot p_\mu) (U(p_0) \cdot p_\nu) = - p_0^2 f^2(p_0) + p_i p_i g^2(p_0) = - m_0^2,
\end{equation}
where the line element is given as
\begin{equation}
\label{length}
ds^2=-\frac{1}{f^2}dt^2 +\frac{1}{g^2}dx^idx^i
\end{equation}
 in the energy-independent coordinates,
 and the rainbow functions $f$ and $g$ depend
 on the energy of test particles.
Note that $U(p_0) \cdot p_\mu$ can be replaced by $\tilde{p}_\mu$
such that the line element can be written in the form of the energy-dependent fashion.
In particular, if we assume that the commutation relations are defined as
\begin{equation}
[x^\mu, \tilde{p}_\nu] = \delta^\mu_\nu,
\end{equation}
then one can express $x^\mu = \partial_{\tilde{p}_\mu}$.
As a matter of fact, these are equivalent
to the expression in Ref.~\cite{Cortes:2004qn}
where the spacetime coordinates are the generators for translations in the auxiliary
linearly transforming energy-momentum variables $\tilde{p}_\nu$.
So the energy-dependent coordinates are related to the energy-independent ones
as follows
\begin{equation}
\label{coordinate relation}
\tilde{x}^0 = f^{-1} x^0,~~ \tilde{x}^i = g^{-1} x^i
\end{equation}
from the flat metric~\eqref{length}, while
$\tilde{p}_0 = p_0 f$ and $\tilde{p}_i = p_i g$ are read off
from the dispersion relation~\eqref{mdr}.
Now, the commutation relations can be written by the energy-independent variables
by using the chain rules,
\begin{align}
  \label{msup:x0}
  x^0 &= (\partial_{\tilde{p}_0} p_0) \partial_{p_0} + (\partial_{\tilde{p}_0} p_i) \partial_{p_i} = \frac{1}{(f + p_0 f')g} \left[ (g + p_0 g') \partial_{p_0} - g'  p_\alpha \partial_{p_\alpha} \right], \\
  \label{msup:xi}
  x^i &= (\partial_{\tilde{p}_i} p_0) \partial_{p_0} + (\partial_{\tilde{p}_i} p_j) \partial_{p_j} = \frac{1}{g} \partial_{p_i},
\end{align}
where the prime denotes the derivative with respect to $p_0$.
Using Eqs.~\eqref{msup:x0} and \eqref{msup:xi},
the commutation relations between the original variables are obtained as
\begin{align}
  \label{msup:00}
  [x^0, p_0] &= \frac{1}{f + p_0 f'}, \\
  \label{msup:0i}
  [x^0, p_i] &= - \frac{p_i g'}{(f + p_0 f')g}, \\
  \label{msup:ij}
  [x^i, p_j] &= \delta^i_j \frac{1}{g},
\end{align}
where they are coincident with the results
in Refs.~\cite{Magueijo:2002am,Cortes:2004qn}.
From Eq.~\eqref{msup:ij},
the uncertainty relation between the
position and the canonical momentum becomes
\begin{equation}
  \label{msup}
  \Delta x \Delta p \ge \frac{1}{2g}.
\end{equation}

Let us now consider the metric of the rainbow Schwarzschild black hole
in order to apply the above uncertainty relation to the
calculation of Hawking temperature.
The metric is given as~\cite{Magueijo:2002xx},
\begin{equation}
  \label{rSBH}
  ds^2 = - \left( 1 - \frac{2GM}{r^2} \right) \frac{dt^2}{f^2} + \left( 1 - \frac{2GM}{r^2} \right)^{-1} \frac{dr^2}{g^2} + \frac{r^2}{g^2} d\Omega^2,
\end{equation}
where the event horizon is $r_H=2GM$ and the metric is
reduced to the flat rainbow metric~\eqref{length} for $M=0$.
The Hawking temperature in Gravity's Rainbow
measured by the asymptotic observer at infinity
can be derived from the heuristic method~\cite{Adler:2001vs}.
From the dispersion relation~\eqref{mdr},
one can express the energy for the massless particles as
$E = (g/f)p= (g/f)\Delta p$, and
it becomes $E = (g/f) (2g \Delta x)^{-1} = (2f \Delta x)^{-1}$
by employing the uncertainty principle~\eqref{msup}.
Thus the rainbow Hawking temperature is written as
\begin{equation}
\label{ht1}
  T_H = \frac{1}{8\pi GM f}
\end{equation}
with the calibration factor $2\pi$, where $\Delta x \sim 2GM$.
By the way, the Hawking temperature can also be derived from the calculation of
the surface gravity of the
metric~\eqref{rSBH}~\cite{Ali:2014xqa,Gim:2014ira}, which is given as
\begin{equation}
\label{genuine ht}
  T_H = \frac{1}{8\pi GM} \frac{g}{f}.
\end{equation}
Note that the temperature~\eqref{ht1} is different from the
temperature~\eqref{genuine ht} in that
the former case is independent of the rainbow function $g$.

The above calculation will be repeated by assuming
a different uncertainty relation from Eq.~\eqref{msup}.
For this purpose, let us assume commutation relations defined as
\begin{equation}
\label{xp}
[\tilde{x}^\mu, \tilde{p}_\nu] = \delta^\mu_\nu.
\end{equation}
From the
inverse representation of Eq.~\eqref{coordinate relation}, the coordinates are expressed by
\begin{align}
  x^0 &= f \tilde{x}^0 = \frac{1}{1 + p_0 (f'/f)} \left[ (1 + p_0 (g'/g)) \partial_{p_0} - (g'/g)
   p_\alpha \partial_{p_\alpha}\right], \\
  x^i &= g \tilde{x}^i = \partial_{p_i},
\end{align}
where the energy-dependent coordinates are represented as
$\tilde{x}^\mu = \partial_{\tilde{p}_\mu}$ from Eq.~\eqref{xp}.
So the nontrivial commutation relations between the spacetime coordinates
are calculated as
\begin{equation}
\label{1}
  [x^0, x^i] = \frac{(g'/g)}{1 + p_0 (f'/f)} x^i,
\end{equation}
and the other nonvanishing commutation relations become
\begin{align}
  [x^0, p_0] &= \frac{1}{1 + p_0 (f'/f)}, \label{2} \\
  [x^0, p_i] &= - \frac{(g'/g)}{1 + p_0 (f'/f)} p_i, \label{3} \\
  [x^i, p_j] &= \delta^i_j. \label{4}
\end{align}
Note that the Heisenberg uncertainty relation
between the coordinate and the momentum is still valid
even in the rainbow regime if we require
the standard commutation relations in the energy dependent frame.
So the Hawking temperature of the rainbow Schwarzschild black hole is now obtained
by using the modified dispersion relation~\eqref{mdr} and the standard uncertainty relation from Eq.~\eqref{4} as
\begin{equation}
  T_H = \frac{1}{8\pi GM} \frac{g}{f},
\end{equation}
where we used the above mentioned heuristic method by identifying
$\Delta x \sim 2GM$, $E=(g/f) \Delta p$.
Note that this is exactly the same as the temperature
calculated from the surface gravity~\eqref{genuine ht} in Refs.~\cite{Ali:2014xqa,Gim:2014ira}.



In conclusion,
we obtained the commutation relations~\eqref{1}--\eqref{4} with
the arbitrary rainbow functions.
Exceptionally, Eq.~\eqref{4} is independent of the rainbow function, so that
the uncertainty relation between the coordinate and the momentum respects
the usual Heisenberg uncertainty relation which provides
the compatible calculation of the temperature with
the Hawking temperature from the surface gravity.

\acknowledgments
This work was supported by the National Research Foundation of Korea(NRF) grant funded by the Korea government(MSIP) (2014R1A2A1A11049571).


\end{document}